\documentclass[onecolumn]{aamas2016}
\usepackage[T1]{fontenc}
\usepackage[latin9]{inputenc}
\usepackage{pmboxdraw}
\usepackage{amsmath}
\usepackage{cancel}
\usepackage{graphicx}
\usepackage{esint}
\PassOptionsToPackage{normalem}{ulem}
\usepackage{ulem}

\makeatletter

\newcommand{\lyxmathsym}[1]{\ifmmode\begingroup\def\b@ld{bold}
  \text{\ifx\math@version\b@ld\bfseries\fi#1}\endgroup\else#1\fi}

\usepackage{url}

\usepackage{subfigure}

\usepackage{graphicx}
\usepackage{animate}
\allowdisplaybreaks

\usepackage{balance}

\makeatother

\begin{document}

\title{Variational Inference with Agent-Based Models}

\author{\alignauthor Wen Dong\\
\affaddr{Department of Computer Science and Engineering \&}\\
\affaddr{Institute of Sustainable Transportation and Logistics}\\
\affaddr{State University of New York at Buffalo}\\
\email{wendong@buffalo.edu}}
\maketitle
\begin{abstract}
In this paper, we develop a variational method to track and make predictions
about a real-world system from continuous imperfect observations about
this system, using an agent-based model that describes the system
dynamics. By combining the power of big data with the power of model-thinking
in the stochastic process framework, we can make many valuable predictions.
We show how to track the spread of an epidemic at the individual level
and how to make short-term predictions about traffic congestion. This
method points to a new way to bring together modelers and data miners
by turning the real world into a living lab.
\end{abstract}

\section{Introduction}

Agent-based modeling has been employed by researchers in many disciplines
to specify the elements of a complex system and their interactions,
to check their understandings of a system, to conduct thought experiments
and to inform design and analysis \cite{Epstein07GenerativeSocialScience,Wilkinson_Stochastic_modelling_2006,borshchev2013big,Nigel07ABM}.
With the availability of big data in recent years \cite{de2014d4d,leetaru2013gdelt,dong2011modeling},
we hope to track and make predictions about a real world system from
the data that represent the continuous observations of this system
and an agent-based model that specifies how the system evolves, and
consequently to turn our world into a living lab. In this paper, we
identify the agent-based model as a discrete-event Markov process,
and develop a variational inference method that searches the latent
state trajectories of the elements of the system in the probability
space that are most compatible with the noisy observations by minimizing
the Bethe variational principle \cite{bethe1935statistical}.

Data have traditionally been used by agent-based modelers to calibrate
model parameters, drive model execution, and validate the model. The
inquiry in this paper is instead about how continuous imperfect observations
about a real-world system can help us make inferences about the system
here and now. Instead of simulating traffic jams at rush hours using
road network data and agent trips synthesized from census data \cite{matsim,smith1995transims},
we are more interested in predicting whether today's traffic jams
will be formed earlier or last longer from the trajectories of probe
vehicles, and how the knowledge about future traffic states will help
drivers use the road network more efficiently. Instead of constructing
the S-shaped curve of an infectious population from simulation \cite{Castellano09SocialDynamics,liggett2012interacting},
we are more interested in who got a sniffle from his dynamic social
network and how we can prevent epidemics from further spreading \cite{DaTr07,DBLP:conf/uai/DongPH12,fan2015hierarchical}.
Instead of showing the emergence of cities and roads from how people
explore and exploit resources \cite{batty2007cities,waddell2002urbansim,forrester1969urban},
We are more interested in identifying poverty and extracting census
information from how people make phone calls \cite{blondel2015survey,de2014d4d,pokhriyal2015virtual}.
Predictions with an agent-based model about real-world data are interpretable
in terms of how agents interact with one another and change states,
and are amenable to reason regarding non-recurrent scenarios. This
transparency about the predictions is lacking in non-parametric approaches.

Our approach is to identify an agent-based simulator as a Markov process,
and to search in the probability space specified by the simulator
for agent behaviors and interactions that best match the continuous
observations about our real-world system. The key observation behind
this approach is that an agent-based simulator generates different
sample paths with different probabilities \textemdash{} it therefore
defines a stochastic process with a probability measure assigned to
the space of the sample paths that describe the interactions among
the elements of the system. In this stochastic process, the system
state as a function of time is composed of the states of its elements.
This stochastic process is driven by a number of events that change
the system state and happen with event rates that are functions of
the current system state. A sample path of the stochastic process
is defined by a sequence of events and the corresponding times when
those events happened, from which we can unambiguously recover the
system state as a function of time. An agent-based simulator therefore
iteratively samples the next event according to event rates then changes
the world state according to the sampled state starting from the initial
state, until the required amount of simulated time has passed.

To find out the maximum likelihood probability distribution of the
system state $X_{t}$ for $t_{1}\le t\le t_{2}$ from observed system
state $x_{t_{1}}$ at time $t_{1}$ and $x_{t_{2}}$ at time $t_{2}$,
we follow the forward-backward algorithm: we first let the probability
mass diffuse from $x_{t_{1}}$ in the forward step according to how
this system evolves from $t_{1}$ to $t_{2}$, and then we iteratively
trace backwards from $t_{2}$ to $t_{1}$ in the backward step how
the probability mass ended at system state $x_{t_{2}}$ instead of
another state. After the forward step and the backward step, we get
the probability distribution of system state $X_{t}$ for $t_{1}\le t\le t_{2}$
conditioned on both of its states $x_{t_{1}}$ and $x_{t_{2}}$.

The challenge in making probabilistic inferences about an agent-based
model is that we have to deal with an exploding state space --- for
just a simple task of tracking the binary states of 50 agents, we
must cope with $2^{50}$ combinatorial states because the agents interact
with one another. And, of course, a real-world system is much larger.
To cope with this exploding state space, we use mean field approximation:
the probabilistic evolution of an agent state is determined by the
mean field (average) effect of the states of the other agents. The
variational framework for making inferences about stochastic processes
was developed in the field of machine learning \cite{DBLP:journals/ftml/WainwrightJ08}
as minimizing Bethe variational principle \cite{bethe1935statistical}
with applications to expectation propagation \cite{DBLP:conf/uai/Minka01,DBLP:conf/uai/HeskesZ02}
and loopy belief propagation \cite{murphy1999loopy}.

This paper therefore advocates that we should combine the power of
big data and the power of model-thinking in the stochastic process
framework. Agent-based modeling is a physicist's approach for modeling
human societies when data are unavailable and experiments impossible
\cite{Epstein07GenerativeSocialScience}, and we believe that big
data will transform agent-based modeling from speculation into a physical
science. This paper also offers a solution that fits to big time-series
data any agent-based model defined by a production rule system based
on mean-field approximation. Hence, this system brings together modelers
and data miners. We have benchmarked our solution on systems of hundreds
of agents, and our benchmarking gives meaningful results.

The rest of this paper is organized as follows. In Section \ref{sec:ProbabilisticProductionSystem}
we introduce a probabilistic production (rule) system to describe
the microscopic dynamics of a generative model, and identify the production
system as a stochastic process. In Section \ref{sec:Inference} we
derive a mean-field solution to the generative model under the constraint
of data. In other words, given a simulator and noisy observational
data about a process generated by the simulator logic, our algorithm
infers the probabilities on a per-agent basis of all possible outcomes.
In Section \ref{sec:examples} we give examples and benchmark this
algorithm against other algorithms. We summarize what we have accomplished
and offer our speculation about big data in Section \ref{sec:Conclusions}.

\section{Stochastic Process Induced by Agent-Based Models}

\label{sec:ProbabilisticProductionSystem}In this section, we introduce
the stochastic kinetic model described by the Gillespie algorithm
\cite{GiDa07} to make inferences about social dynamics from information
about individuals in the social system. A ``stochastic kinetic model''
is a chemist's way of describing the temporal evolution of a system
with $M$ agent species driven by $V$ events (or chemical reactions)
parameterized by rate constants $c=(c_{1},\dots,c_{V})$. At any specific
time $t$, the populations of the species are $x_{t}=(x_{t}^{(1)},\dots,x_{t}^{(M)})$.
An event $v$ happens with rate $h_{v}(x_{t},c_{v})$, changing the
populations by $\Delta_{v}$. The $V$ events are mutually independent.\\
\\
\\
\rule{1\columnwidth}{1pt}

\textbf{\uline{Gillespie algorithm\hspace*{\fill}}} 
\begin{enumerate}
\item Initialize the system at time $t=0$ with rate constants $c_{1},\dots,c_{V}$
and initialize the populations of the species as $x^{(1)},\dots x^{(M)}$. 
\item Simulate the time $\tau$ to the next event according to exponential
distribution $\tau\sim\mbox{Exponential}(h_{0}(x,c)=\sum_{v=1}^{V}h_{v}(x,c_{v}))$.
\item Simulate the event $v$ according to categorical distribution $v\sim\mbox{Categorical}(\frac{h_{1}}{h_{0}},\dots,\frac{h_{V}}{h_{0}})$.
\item Update and output time $t\leftarrow t+\tau$ and populations $x\leftarrow x+\Delta_{v}$. 
\item Repeat steps 2-5 until the termination condition is satisfied. 
\end{enumerate}
\rule{1\columnwidth}{1pt}

The stochastic kinetic model specified by the Gillespie algorithm
assigns a probability measure to a sample path of the system induced
by a sequence of events $v_{1},\dots,v_{n}$, happening between time
$0$ and time $T$, $0=t_{0}<t_{1}<\dots<t_{n}<T$, which is 
\begin{align}
 & P(v_{1},\dots,v_{n},t_{1},\dots,t_{n},x)\label{eq:CTSKM}\\
= & \prod_{i=1}^{n}h_{v_{i}}(x_{t_{i-1}},c_{v_{i}})\exp\left(-\sum_{i=1}^{n}h_{0}(x_{t_{i-1}},c)(t_{i}-t_{i-1})\right)\nonumber \\
= & \prod_{i=1}^{n}h_{v_{i}}(x_{t_{i-1}},c_{v_{i}})\exp\left(-\int d\ t\ h_{0}(x_{t},c)\right).\nonumber 
\end{align}

The stochastic kinetic model was designed to explain the macroscopic
properties of a system via the microscopic interactions among particles
in the system \cite{Wilkinson_Stochastic_modelling_2006}. If we are
able to observe particles, we expect not only to improve our estimation
of the system properties but also to make inferences about the particles.
When this model is applied to a social system, the particles are the
individuals in the system, and the capability to make inferences about
these particles becomes even more important. An event in the Gillespie
algorithm looks like the following:
\begin{align}
 & {\scriptstyle \alpha_{v}^{(1)}X^{(1)}+\cdots+\alpha_{v}^{(M)}X^{(M)}\to\beta_{v}^{(1)}X^{(1)}+\cdots+\beta_{v}^{(M)}X^{(M)}},\label{eq:reaction}\\
 & {\scriptstyle h_{v}(x,c_{k})=c_{v}g_{v}(x)=c_{v}\prod_{m=1}^{M}g_{v}^{(m)}(x^{(m)})=c_{v}\prod_{m=1}^{M}(x^{(m)})^{\alpha_{v}^{(m)}}},\label{eq:event}\\
 & \Delta_{v}=(\beta_{v}^{(1)}-\alpha_{v}^{(1)},\cdots,\beta_{v}^{(M)}-\alpha_{v}^{(M)}).\nonumber 
\end{align}
When $\alpha_{v}^{(1)}$ individuals of species $1$, $\alpha_{v}^{(2)}$
individuals of species $2$ ... meet, they trigger event $v$ with
rate constant $c_{v}$, which result in $\beta_{v}^{(1)}$ individuals
of species $1$, $\beta_{v}^{(2)}$ individuals of species $2$, and
so on. The rate $h_{v}(x,c_{v})$ for this event to happen is rate
constant $c_{v}$ times a total of $\prod_{m=1}^{M}\left(x^{(m)}\right)^{\alpha_{v}^{(m)}}$
different ways for the individuals to meet. 

When the components of the agent-based model $X^{(1)}$, $\cdots$,
$X^{(M)}$ lose meaning as the populations of agent species, we can
find the probability distribution of a component state among a finite
partition of the component's state space. When the event rates cannot
be expressed as the multiplications of component contributions in
the form of Eq. \ref{eq:event}, we take Taylor expansions of the
event rates around the mean value of the system state 
\begin{eqnarray*}
h_{v}(x,c_{v}) & = & \sum_{|\alpha|\ge0}\frac{(x-\mathbf{E}x)^{\alpha}}{\alpha!}\partial^{\alpha}h_{v}(\mathbf{E}x,c_{v}),
\end{eqnarray*}
with each term being in the form of Eq. \ref{eq:event}. $\alpha$
is a multi-index and $\mbox{E}$ is the expectation operator.

Although the stochastic kinetic model is a continuous time model,
we work with a discrete time stochastic model in the rest of this
paper, because our goal is to track stochastic kinetic dynamics from
observations of populations or individuals with countably many computational
steps. There are two ways to turn a continuous stochastic process
into a discrete one, both of which involve Jensen's uniformization/randomization
method \cite{grassman77uniformization}.

The first method for discretizing a continuous time stochastic system
is through approximating the continuous time process with a discrete
time process on a countable set of equally spaced time points $0,\tau,2\tau,\dots$,
with a time interval so small that the probability of more than one
event happening in the interval $\tau$ is negligible. This approximation
works because the state transition kernel from time $0$ to time $\tau$
is $p(x_{0}\to x_{\tau})=\sum_{n=0}^{\infty}\left(I+\frac{Q}{\gamma}\right)^{n}\exp(-\gamma\tau)\frac{(\gamma\tau)^{n}}{n!}$
according to the uniformization method, where $\gamma$ is a uniformization
rate, $I$ is the identity matrix and $Q$ is the infinitesimal generator
defined by $h_{k},k=1,\dots,V$. With $\gamma\to\infty$ and $\gamma\tau=1$,
we get a first-order approximation of the state transition kernel
$I+Q\cdot\tau$.

Specifically, let $v_{1},\dots,v_{T}$ be a sequence of events in
the discrete time stochastic kinetic system, $x_{1},\dots,x_{T}$
be a sequence of states (populations of species), and $y_{1},\dots,y_{T}$
be a set of observations about the populations. Our goal is to make
inferences about $\{v_{t},x_{t}:t=1,\dots T\}$ from $\{y_{t}:t=1,\dots,T\}$
according to the following probability measure, where indicator function
$1(x_{t}-x_{t-1}=\Delta_{v_{t}})$ is 1 if the previous state is $x_{t-1}$
and the current state is $x_{t}=x_{t-1}+\Delta_{v_{t}}$, and 0 otherwise.
\begin{align}
 & P\left(v_{1,\dots,T},x_{1,\dots,T},y_{1,\dots,T}\right)=\prod_{t=1}^{T}P(x_{t},y_{t},v_{t}|x_{t-1}),\label{eq:DTSKM}\\
 & \mbox{where }P(x_{t},y_{t},v_{t}|x_{t-1})\nonumber \\
 & \hspace*{5em}=P(v_{t}|x_{t-1})1(x_{t}-x_{t-1}=\Delta_{v_{t}})P(y_{t}|x_{t}),\label{eq:DTSKMTransition}\\
 & \mbox{and }\hspace*{1em}P(v_{t}|x_{t-1})=\begin{cases}
c_{k}\tau\cdot g_{k}\left(x_{t-1}\right) & \mbox{if }v_{t}=k\\
1-\sum_{j}c_{j}\tau g_{j}\left(x_{t-1}\right) & \mbox{if }v_{t}=\emptyset
\end{cases}.\label{eq:DTSKMP}
\end{align}

The second way to discretize a continuous time stochastic system is
by introducing a uniformization rate $\gamma$ that is faster than
all event rates in $Q$ and inspecting a discrete time Markov chain
defined by the state transition matrix $I+\frac{Q}{\lambda}$, with
the transitions happening at time $t_{1},t_{2},\dots,$ sampled according
to a uniform Poisson process with rate $\gamma$. This works because
according to the uniformization method the uniformised continuous
time process has the same probability measure as the original process.

We employ a stochastic kinetic model to simplify the state space transition
kernel for several reasons. First, the stochastic kinetic model already
successfully describes the time evolution of reaction systems in many
areas, including chemistry and cell biology \cite{ArRo98,GiDa07}.
It is therefore a more natural model for describing and tracking the
spatio-temporal process driven by events. Second, the event based
transition kernel is more general and flexible--we can define the
number of events based on the complexity of real transitions.

\section{Making Inferences with an \protect \\
Agent-Based Model}

\label{sec:Inference}In this section, we derive a mean-field solution
to infer the probabilities on a per-agent basis of all possible paths
of system evolution, given a simulator and noisy observational data
about this system generated by the simulator logic.

\subsection{Variational Inference}

Recall the forward-backward algorithm to make inferences with a state
space model \cite{rabiner1989tutorial}. Let $X_{t}$ be the hidden
states and $y_{t}$ be the observations of a discrete-time state-space
model (Kalman filter and hidden Markov model) identified by a transition
probability $P(X_{t+1}|X_{t})$ and an observation model $P(Y_{t}|X_{t})$,
where $t=1,\cdots,T$. The forward-backward algorithm for making inferences
about hidden states $X_{t}$ from observations $y_{t}$ is comprised
of a forward/filtering sweep to compute the forward statistics $\alpha(x_{t})=P(x_{t}|y_{1},\cdots,y_{T})$
and a backward/smoothing sweep to estimate the one-slice statistics
$\gamma(y_{t})=P(x_{t}\lyxmathsym{\textSFxi}y_{1},\cdots,y_{T})$.
From the forward statistics and the one-slice statistics we can extract
the backward statistics $\beta(x_{t})=\gamma(x_{t})/\alpha(x_{t})$
and the two-slice statistics $\xi(x_{t},x_{t+1})=\alpha(x_{t})P(y_{t+1},x_{t+1}\lyxmathsym{\textSFxi}x_{t})\beta(x_{t+1})P(y_{t+1}|y_{1,\cdots,t})$.
\\
Here we follow the tradition, use upper case letters for random variables
and use lower case letters for the values of random variables.

The challenge with making inferences about a non-trivial agent-based
model is that we have to search in a formidable state space --- $X_{t}=(X_{t}^{(1)},X_{t}^{(2)},\cdots,X_{t}^{(M)})$,
where the superscripts $1,\cdots,M$ represent the states of the interacting
elements of the system. We therefore estimate the state distributions
of the hidden states in an amenable state space with mean field approximation
$\gamma_{t}(x_{t})=\prod_{m}\gamma_{t}^{(m)}(x_{t}^{(m)})$:
\begin{align}
 & \mbox{minimize over }\xi_{t}(x_{t-1,t},v_{t}):\nonumber \\
 & \sum\limits _{t;x_{t-1,t};v_{t}\hidewidth}\xi_{t}(x_{t-1,t},v_{t})\log\frac{\xi_{t}(x_{t-1,t},v_{t})}{P(x_{t},v_{t-1},y_{t}|x_{t-1})}\label{eq:KL}\\
 & \hspace*{5em}-\sum\limits _{t;x_{t}}\prod\limits _{m}\gamma_{t}^{(m)}(x_{t}^{(m)})\log\prod\limits _{m}\gamma_{t}^{(m)}(x_{t}^{(m)})\nonumber \\
 & \mbox{subject to: }\nonumber \\
 & \sum_{v_{t};x_{t-1,t}:\mbox{ fixed }x_{t}^{(m)}\hidewidth}\xi_{t}(x_{t-1},x_{t},v_{t})=\gamma_{t}^{(m)}(x_{t}^{(m)})\mbox{, for all }t,m,x_{t}^{(m)},\\
 & \sum_{v_{t};x_{t-1,t}:\mbox{ fixed }x_{t-1}^{(m)}\hidewidth}\xi_{t}(x_{t-1},x_{t},v_{t})=\gamma_{t-1}^{(m)}(x_{t-1}^{(m)})\mbox{, for all }t,m,x_{t-1}^{(m)},\\
 & \sum_{x_{t}:\mbox{ fixed }x_{t}^{(m)}\hidewidth}\gamma_{t}^{(m)}(x_{t}^{(m)})=1\mbox{, for all }t,m,x_{t}^{(m)}.
\end{align}

We apply the method of Lagrange multipliers to solve this optimization
problem, which begins with forming the Lagrange function to be optimized:

\begin{align}
 & \sum\limits _{t;x_{t-1,t};v_{t}\hidewidth}\xi_{t}(x_{t-1,t},v_{t})\log\frac{\xi_{t}(x_{t-1,t},v_{t})}{P(x_{t},v_{t-1},y_{t}|x_{t-1})}\label{eq:Lagrange}\\
 & \hspace*{5em}-\sum\limits _{t;x_{t}}\prod\limits _{m}\gamma_{t}^{(m)}(x_{t}^{(m)})\log\prod\limits _{m}\gamma_{t}^{(m)}(x_{t}^{(m)})\nonumber \\
 & +\sum_{\hidewidth t;m;x_{t}^{(m)}}\beta_{t,x_{t}^{(m)}}^{(m)}(\sum_{v_{t};x_{t-1,t}:\mbox{ fixed }x_{t}^{(m)}\hidewidth}\xi_{t}(x_{t-1},x_{t},v_{t})-\gamma_{t}^{(m)}(x_{t}^{(m)}))\nonumber \\
 & +\sum_{\hidewidth t;m;x_{t-1}^{(m)}}\alpha_{t-1,x_{t-1}^{(m)}}^{(m)}(\sum_{v_{t};x_{t-1,t}:\mbox{ fixed }x_{t-1}^{(m)}\hidewidth}\xi_{t}(x_{t-1},x_{t},v_{t})-\gamma_{t-1}^{(m)}(x_{t-1}^{(m)})).\nonumber 
\end{align}

Taking the derivative of the expression involving Lagrange multipliers
over $\xi_{t}(x_{t-1},x_{t},v_{t})$ and $\gamma_{t}^{(m)}(x_{t}^{(m)})$,
we see that ${\scriptstyle \alpha_{t}^{(m)}(x_{t}^{(m)})=\exp(\sum_{i}\alpha_{t,i}^{(m)}\cdot1(x_{t}^{(m)}=i))}$
is associated with the marginalized forward probabilities, ${\scriptstyle \beta_{t}^{(m)}(x_{t}^{(m)})=\exp(\sum_{i}\beta_{t,i}^{(m)}\cdot1(x_{t}^{(m)}=i))}$
is associated with the marginalized backward probabilities, with $\gamma_{t}^{(m)}(x_{t}^{(m)})=\alpha_{t}^{(m)}(x_{t}^{(m)})\beta_{t}^{(m)}(x_{t}^{(m)})$.
The dual optimization problem is to find the marginal forward statistics
$\alpha_{t}^{(m)}(x_{t}^{(m)})$ and the marginal backward statistics
$\beta_{t}^{(m)}(x_{t}^{(m)})$ to maximize the approximate partition
function given by Eq. \ref{eq:SKMEvidence}, and the solution is the
fixed point of Eq. \ref{eq:SKM2Slice}, where normalization constant
$Z_{t}=P(y_{t}|y_{1,\cdots,t-1})$:

\begin{align}
 & {\scriptstyle \log P(y_{1,\cdots,T})=\sum\limits _{t}\log\sum\limits _{x_{t-1,t}}\prod\limits _{m}\alpha_{t-1}^{(m)}(x_{t-1}^{(m)})P(x_{t},v_{t}|x_{t-1})\prod\limits _{m}\beta_{t}^{(m)}(x_{t}^{(m)})}\label{eq:SKMEvidence}\\
 & \xi_{t}(x_{t-1},x_{t},v_{t})=\frac{1}{Z_{t}}P(x_{t},v_{t}|x_{t-1})\label{eq:SKM2Slice}\\
 & \hspace{2em}\cdot\prod_{m}\alpha_{t-1}^{(m)}\left(x_{t-1}^{(m)}\right)\cdot\prod_{m}P\left(y_{t}^{(m)}|x_{t}^{(m)}\right)\prod_{m}\beta_{t}^{(m)}\left(x_{t}^{(m)}\right),\nonumber \\
 & \mbox{where }P(x_{t},v_{t}|x_{t-1})=\label{eq:SKMEventRate}\\
 & \hspace*{1em}\begin{cases}
{\scriptstyle c_{k}\tau\prod_{m}g_{k}^{(m)}\left(x_{t-1}^{(m)}\right)\cdot\prod_{m}1(x_{t}^{(m)}-x_{t-1}^{(m)}=\Delta_{k}^{(m)})} & v_{t}=k\ne\emptyset\\
{\scriptstyle (1-\sum_{k}c_{k}\tau\prod_{m}g_{k}^{(m)}(x_{t-1}^{(m)}))\cdot\prod_{m}1(x_{t}^{(m)}-x_{t-1}^{(m)}=0)} & v_{t}=\emptyset.
\end{cases}\nonumber 
\end{align}

Marginalizing Eq. \ref{eq:SKM2Slice} over all chains $X_{t}^{(m')}$
for $m'\ne m$ and $t$, we find that the solution to the Bethe variational
principle is the mean field approximation of the original dynamics
with marginal two-slice statistic given by Eq. \ref{eq:marginal2Slice}.
From the mean field approximation, $\alpha_{t,i}^{(m)}$ and $\beta_{t,i}^{(m)}$
can be solved by forward-backward algorithm and fixed point iteration.

\begin{align}
 & \xi_{t}(x_{t-1}^{(m)},x_{t}^{(m)},v_{t})=\frac{1}{Z_{t}}P(x_{t}^{(m)},v_{t}|x_{t-1}^{(m)})\label{eq:marginal2Slice}\\
 & \hspace{7em}\cdot\alpha_{t-1}^{(m)}(x_{t-1}^{(m)})\cdot P(y_{t}^{(m)}|x_{t}^{(m)})\beta_{t}^{(m)}(x_{t}^{(m)}),\nonumber \\
 & \mbox{where }P(x_{t}^{(m)},v_{t}|x_{t-1}^{(m)})\propto\label{eq:marginalEventRate}\\
 & \hspace*{1em}\begin{cases}
{\scriptstyle c_{k}\tau g_{k}^{(m)}(x_{t-1}^{(m)})\prod\limits _{m'\ne m}\tilde{g}_{k,t-1}^{(m')}\cdot1(x_{t}^{(m)}-x_{t-1}^{(m)}=\Delta_{k}^{(m)})} & {\scriptstyle v_{t}^{(m)}=k\ne\emptyset}\\
{\scriptstyle \left(1-\sum\limits _{k}c_{k}\tau g_{k}^{(m)}(x_{t-1}^{(m)})\prod\limits _{m'\ne m}\hat{g}_{k,t-1}^{(m')}\right)1(x_{t}^{(m)}-x_{t-1}^{(m)}=0)} & v_{t}^{(m)}=\emptyset,
\end{cases}\nonumber \\
 & \tilde{g}_{k,t-1}^{(m')}=\frac{\sum\limits _{x_{t}^{(m')}-x_{t-1}^{(m')}\equiv\Delta_{k}^{(m')}\hidewidth}{\scriptstyle \alpha_{t-1}^{(m')}(x_{t-1}^{(m')})P(y_{t}^{(m')}|x_{t}^{(m')})\beta_{t}^{(m')}(x_{t}^{(m')})g_{k}^{(m')}(x_{t-1}^{(m')})}}{\sum\limits _{x_{t}^{(m')}-x_{t-1}^{(m')}\equiv0\hidewidth}\alpha_{t-1}^{(m')}(x_{t-1}^{(m')})P(y_{t}^{(m')}|x_{t}^{(m')})\beta_{t}^{(m')}(x_{t}^{(m')})},\nonumber \\
 & \hat{g}_{k,t-1}^{(m')}=\frac{\sum\limits _{x_{t}^{(m')}-x_{t-1}^{(m')}\equiv0\hidewidth}{\scriptstyle \alpha_{t-1}^{(m')}(x_{t-1}^{(m')})P(y_{t}^{(m')}|x_{t}^{(m')})\beta_{t}^{(m')}(x_{t}^{(m')})g_{k}^{(m')}(x_{t-1}^{(m')})}}{\sum\limits _{x_{t}^{(m')}-x_{t-1}^{(m')}\equiv0\hidewidth}\alpha_{t-1}^{(m')}(x_{t-1}^{(m')})P(y_{t}^{(m')}|x_{t}^{(m')})\beta_{t}^{(m')}(x_{t}^{(m')})},\nonumber \\
 & Z_{t}=\sum_{j}c_{j}\tau\prod_{m}\tilde{g}_{j,t-1}^{(m)}+1-\sum_{j}c_{j}\tau\prod_{m}\hat{g}_{j,t-1}^{(m)}.\nonumber 
\end{align}

The above is a factorized stochastic kinetic model. The marginal two-slice
probability $\xi_{t}(x_{t-1}^{(m)},x_{t}^{(m)},v_{t})$ in Eq. \ref{eq:marginal2Slice}
takes the same form as the coupled two-slice probability $\xi_{t}(x_{t-1},x_{t},v_{t})$
in Eq. \ref{eq:SKM2Slice}. The marginal state transition kernel $P(x_{t}^{(m)},v_{t}|x_{t-1}^{(m)})$
in Eq. \ref{eq:marginalEventRate} consists of choosing an event (or
no event) $v_{t}$ according to event probability $P(v_{t}|x_{t-1}^{(m)})$
and changing the state $x_{t}^{(m)}$ in a deterministic way, similar
to the joint state transition kernel $P(x_{t},v_{t}|x_{t-1})$ in
Eq. \ref{eq:SKMEventRate}, except that we marginalize over all $x^{(m')}$
for $m'\ne m$.

Hence the solution to the above Bethe variational principle through
Legendre-Fenchel transform \cite{rockafellar2015convex} is one in
which the interacting elements of the system evolve their states marginally
according to the average effects of the other elements. As such, instead
of searching the joint probability space of of $(X_{1},\cdots,X_{T})$,
we search the marginal probability spaces of $(X_{1}^{(m)},\cdots,X_{T}^{(m)})$.

\subsection{Graphical Model Representation}

\label{CHMMR}
\begin{figure*}
\centering \subfigure[CHMM]{\label{fig:CHMM}\includegraphics[width=0.32\textwidth]{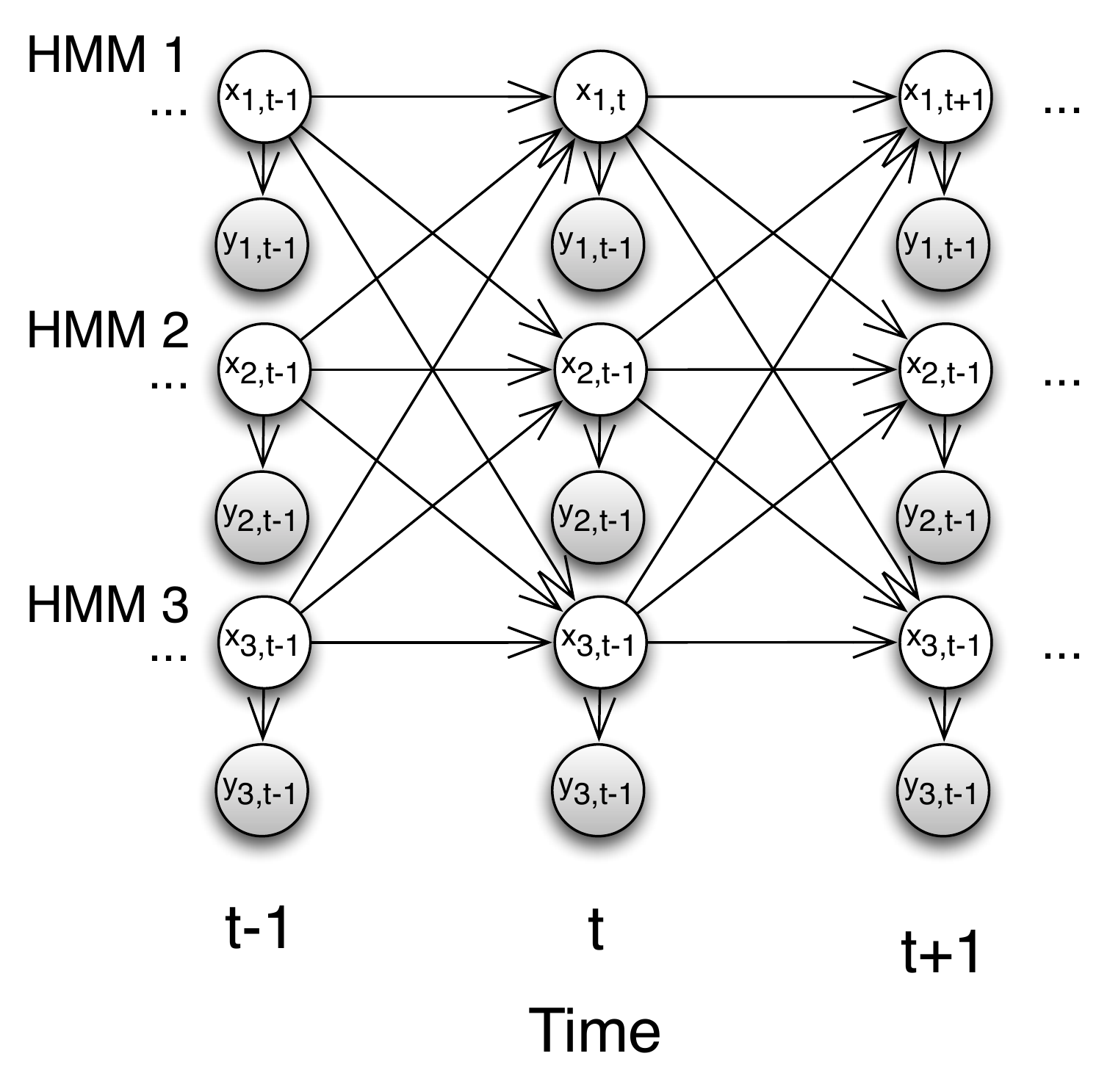}}\subfigure[SKM]{\label{fig:SKM}\includegraphics[clip,width=0.32\textwidth]{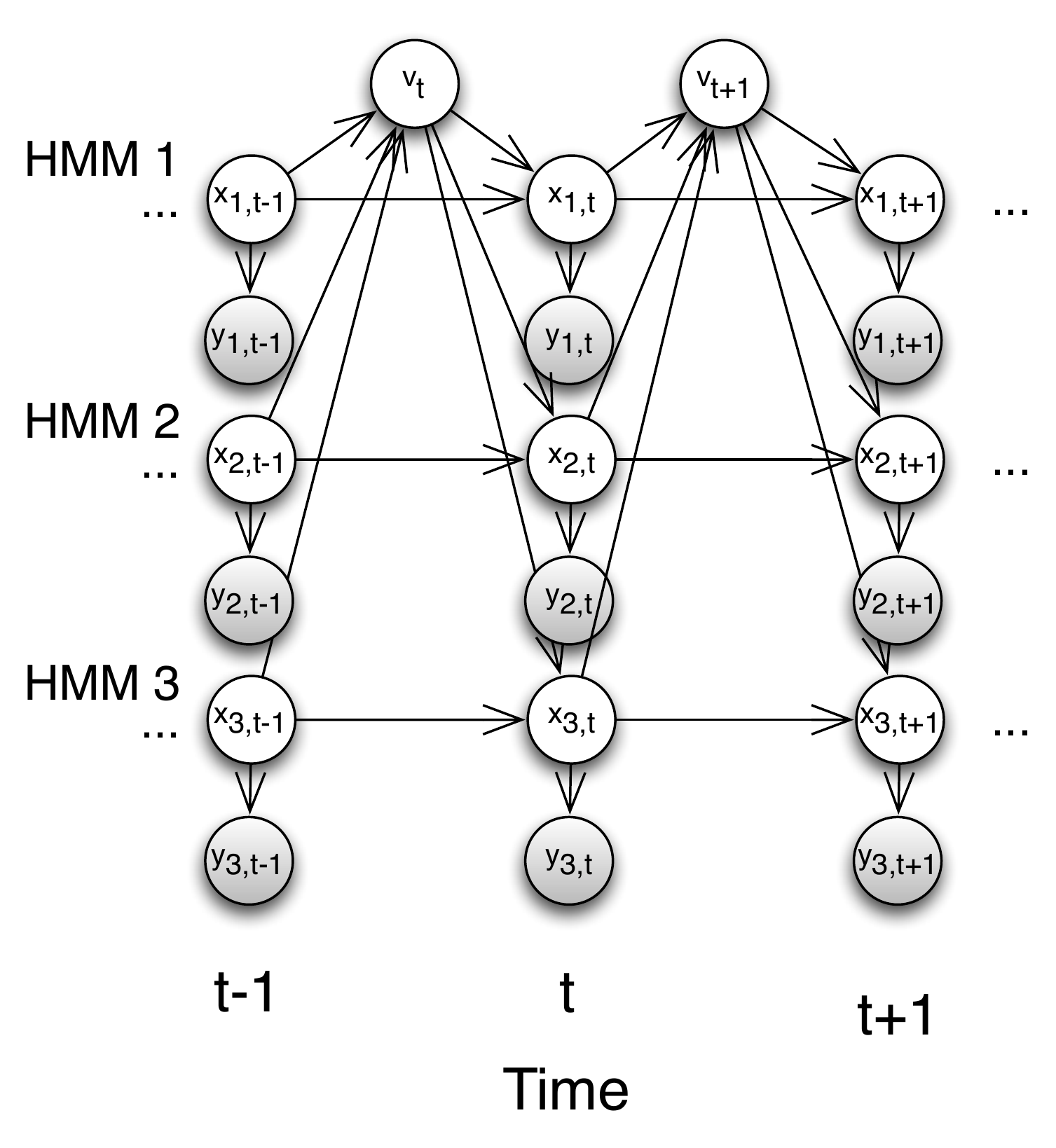}}\subfigure[FSKM]{\label{fig:FSKM}\includegraphics[clip,width=0.32\textwidth]{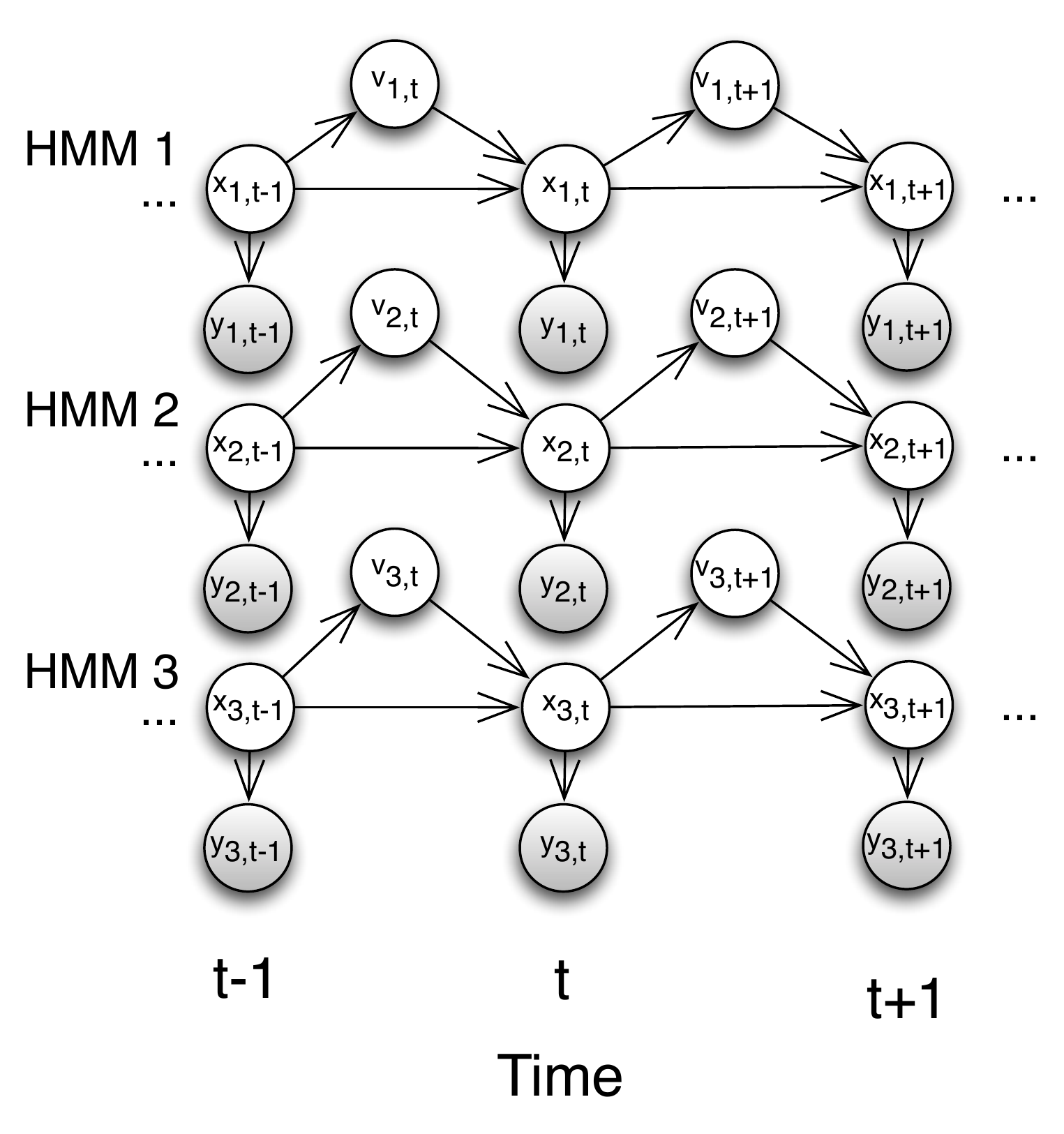}}

\protect\caption{\label{fig:Graphical-model}In comparison of coupled hidden markov
model (a), a stochastic kinetic model (b) decouples inter-chain interactions
with events $v$ and allows factorization (c).}
\end{figure*}

The stochastic kinetic model with its distinct graphical model structure
is more suitable than traditional models for modeling complex interactions
in social dynamics. To illustrate this point, we compare its graphical
model with the coupled hidden Markov model.

A coupled hidden Markov model (CHMM, Figure \ref{fig:CHMM}) combines
a number of conventional hidden Markov models (HMMs) to model the
dynamics of interacting processes \cite{BrOl97,pan2012modeling}.
In CHMM the latent state of HMM at time $t$ depends on the latent
states of \emph{all} HMMs at time $t-1$. Traditional ways to simplify
the state transition kernel include the factorial hidden Markov model
which decouples the inter-chain probability dependence \cite{JoGh97}
and the hidden Markov decision tree which assumes fixed and sparse
inter-chain probability dependence \cite{GhJo97}.

A stochastic kinetic model has a graphical model representation different
from CHMM, as shown in Figure \ref{fig:SKM}. First, we define a set
of stochastic events to summarize the complex interactions and decouple
direct dependencies between nodes. Second, while the system can move
from any state to any other state in a CHMM, in any infinitesimal
time interval no more than one out of $V$ possible events is happening
in a stochastic kinetic model. Third, conditioned on the system state
$x_{t-1}^{(1)},\dots,x_{t-1}^{(M)}$ describing the populations of
species $1,\dots,M$, the latent state at the next time step $x_{t}^{(1)},\dots,x_{t}^{(M)}$
could be dependent. (Consider an event that changes population $m_{1}$
and population $m_{2}$ simultaneously.) In contrast, in a CHMM the
states $x_{t}^{(1)},\dots,x_{t}^{(M)}$ at time $t$ are conditionally
independent given the states $x_{t-1}^{(1)},\dots,x_{t-1}^{(M)}$
at time $t-1$. Thus the inference algorithms of CHMM are not applicable
for modeling the complex interactions in social dynamics driven by
events.

The factorial stochastic kinetic model has a graphical model similar
to the stochastic kinetic model except that it factorizes stochastic
events to individual species (Figure \ref{fig:FSKM}). The new graphical
model further simplifies the inference algorithm.

\subsection{Parameter Learning}

In order to find the rate constants $c_{v}$ in a stochastic kinetic
model and in a factorial stochastic kinetic model, we maximize the
expected log likelihood and the Bethe entropy approximation respectively
over these rate constants.

The likelihood of rate constants $c_{v}$ in a continuous time stochastic
kinetic model with respect to a sample path identified by events $v_{1},\cdots,v_{n}$,
system states to $x_{t_{1}},\cdots,x_{t_{n}}$ and times $t_{1},\cdots,t_{n}$
is given in Eq. \ref{eq:CTSKM}, and the event rates are given in
Eq. \ref{eq:event}. To find the maximum likelihood estimate of the
rate constants, we set the partial derivatives of the log-likelihood
over the rate constants to 0: 
\begin{eqnarray*}
 &  & \log P={\scriptstyle \sum_{i}\log c_{v_{i}}+\log g_{v_{i}}(x_{t_{i-1}})-\sum_{i}\sum_{j}c_{j}g_{j}(x_{t_{i-1}})\cdot(t_{i}-t_{i-1})},\\
 &  & \frac{\partial}{\partial c_{v}}\log P=\frac{\sum_{i}1(v_{i}=v)}{c_{v}}-\sum_{i}g_{v}(x_{t_{i-1}})\cdot(t_{i}-t_{i-1}),\\
 &  & c_{v}=\frac{\sum_{i}1(v_{i}=v)}{\sum_{i}g_{v}(x_{t_{i-1}})(t_{i}-t_{i-1})}=\frac{\sum_{i}1(v_{i}=v)}{\int_{0}^{T}d\ t\ g_{v}(x(t))}.
\end{eqnarray*}
Hence, the maximum likelihood estimate of the rate constants $c_{v}$
is such that the numbers of events that are expected to happen according
to the event rates along the sample path $\int_{0}^{T}dt\ c_{v}g_{v}(x(t))$
match the numbers of events $\sum_{i}1(v_{i}=v)$ that happened in
the sample path. Indicator function $1(v_{i}=v)$ takes value 1 if
$v_{i}=v$ and 0 if $v_{i}\ne v$. 

The likelihood of rate constants $c_{v}$ in a discrete time stochastic
kinetic model with respect to a sample path identified by events $v_{1},\cdots,v_{T}$,
system states $x_{1},\cdots,x_{T}$ and observations $y_{1},\cdots,y_{T}$
is given in Eq. \ref{eq:DTSKM}. To find the maximum likelihood estimate
of the rate constants, we similarly set the partial derivatives of
the log-likelihood over the rate constants to 0: 
\begin{eqnarray*}
 &  & \log P=\sum_{t}\log p(y_{t}|x_{t})+\sum_{t}\log p(v_{t}|x_{t-1})1({\scriptstyle x_{t}-x_{t-1}=\Delta_{v_{t}}})\\
 &  & \frac{\partial}{\partial c_{v}}\log P=\frac{\sum_{t}1(v_{t}=v)}{c_{v}}-\sum_{t}\frac{\tau g_{v}(x_{t-1})1(v_{t}=\emptyset)}{1-\sum_{j}c_{j}\tau g_{j}(x_{t-1})}\\
 &  & c_{v}=\frac{\sum_{t}1(v_{t}=v)}{\sum_{t}\frac{\tau g_{v}(x_{t-1})\cdot1(v_{t}=\emptyset)}{1-\sum_{j}c_{j}\tau g_{j}(x_{t-1})}}\\
 &  & \hspace*{14em}\stackrel{\tau\to0}{\longrightarrow}\frac{\sum_{t}1(v_{t}=v)}{\sum_{t}\tau g_{v}(x_{t-1})}.
\end{eqnarray*}
Hence, the maximum likelihood estimate of the rate constants $c_{v}$
in a discrete-time stochastic kinetic model similarly matches the
numbers of events that are expected to happen according to the event
rates ($\sum_{t}\frac{c_{v}\tau g_{v}(x_{t-1})1(v_{t}=\emptyset)}{1-\sum_{j}c_{j}\tau g_{j}(x_{t-1})}$,
extrapolated from the times of null events) with the number of events
that happened ($\sum_{t}1(v_{t}=v)$). As the interval $\tau$ approaches
0, the probability of a null event ($1-\sum_{j}c_{j}\tau g_{j}(x_{t-1})$)
and the fraction of null events ($\sum_{t=1}^{T}1(v_{t}=\emptyset)/T$)
both approach 1, and the maximum likelihood estimate of the rate constants
in the discrete time stochastic kinetic model approaches the maximum
likelihood estimate in the continuous time stochastic kinetic model.

When the events $v_{1},\cdots,v_{T}$ and the system states $x_{1},\cdots,x_{T}$
are unobserved latent variables, we use the expectation maximization
(EM) algorithm to iteratively search for the rate constants that maximize
the expected log likelihood over the probability distribution of the
latent variables. EM is an iterative method for finding the maximum
likelihood estimate of the parameters in statistical models involving
unobserved latent variables \cite{dempster1977maximum}. It alternates
performing the expectation (E) step, which constructs the expected
log likelihood as a function of the parameters over the probability
distribution of the latent variables using the current estimate for
the parameters, with the maximization (M) step, which computes the
parameters to maximize the expected log likelihood function constructed
in the E step. The estimated parameters are used to determine the
probability distribution of the latent variables in the next E step.

The expected log likelihood over the posterior probability of events
$v_{1},\cdots,v_{T}$ and system states $x_{1},\cdots,x_{T}$ conditioned
on the observations $y_{1},\dots,y_{T}$ takes the form in Eq. \ref{eq:E}.
Maximizing this expected log likelihood by setting its partial derivatives
over the rate constants gives the updated estimate of rate constants
in Eq. \ref{eq:rateConstant}.
\begin{align}
 & \mbox{E}\left(\log P\right)=\sum_{t;x_{t-1,t};v_{t}\hidewidth}{\scriptstyle \xi_{t}(x_{t-1,t},v_{t};c^{\mbox{old}})\cdot\log P(x_{t},y_{t},v_{t}|x_{t-1};c)}\label{eq:E}\\
 & \frac{\partial\mbox{E}(\log P)}{\partial c_{v}}=\sum_{t}\frac{{\scriptstyle \xi(v_{t}=v)}}{c_{v}}-\sum_{\hidewidth t;x_{t-1}}\frac{{\scriptstyle \tau g_{v}(x_{t-1})\xi(x_{t-1},v_{t}=\emptyset)}}{1-\sum_{j}c_{j}\tau g_{j}(x_{t-1})}\stackrel{\mbox{set}}{=}0,\nonumber \\
 & c_{v}=\frac{\sum_{t}\xi(v_{t}=v)}{\sum\limits _{\hidewidth t;x_{t-1}}\frac{\tau g_{v}(x_{t-1})\cdot{\scriptstyle \xi(x_{t-1},v_{t}=\emptyset)}}{1-\sum_{j}c_{j}\tau g_{j}(x_{t-1})}}\label{eq:rateConstant}\\
 & \hspace*{14em}\stackrel{\tau\to0}{\longrightarrow}\frac{\sum\limits _{t}\xi(v_{t}=v)}{\sum\limits _{t;x_{t-1}\hidewidth}\tau{\scriptstyle \gamma(x_{t-1})g_{v}(x_{t-1})}}.\nonumber 
\end{align}

As such, the rate constant $c_{v}$ for event $v$ matches the expected
number of times this event could have happened ($\sum_{t}\frac{c_{v}\tau g_{v}(x_{t-1})\xi(x_{t-1},v_{t}=\emptyset)}{1-\sum_{j}c_{j}\tau g_{j}(x_{t-1})}$)
according to the event rates \\
($c_{j}\tau g_{j}(x_{t-1})$) along the sample path with the expected
number of times the events happened ($\sum_{i}\xi(v_{i}=v)$).

Using Bethe entropy approximation, \\
$\alpha_{t-1}(x_{t-1})=\prod_{m}\alpha_{t-1}^{(m)}(x_{t-1}^{(m)})$
and $\beta_{t}(x_{t})=\prod_{m}\beta_{t}^{(m)}(x_{t}^{(m)})$, and
setting the discretization time interval $\tau$ to be small enough,
the rate constants $c_{v}$ can be updated according to Eq. \ref{eq:marginalRateConst}.
\begin{align}
 & c_{v}=\frac{\sum\limits _{t}\xi(v_{t}=v)}{\sum\limits _{t}\tau\prod_{m}\sum_{x_{t-1}^{(m)}}{\scriptstyle \gamma^{(m)}(x_{t-1}^{(m)})g_{v}^{(m)}(x_{t-1}^{(m)})}},\label{eq:marginalRateConst}\\
 & \mbox{where }\xi(v_{t}=v)=\frac{c_{v}^{\mbox{old}}\prod_{m}\tilde{g}_{v,t-1}^{(m)}}{{\scriptstyle \sum_{j}c_{j}^{\mbox{old}}\tau\prod_{m}\tilde{g}_{j,t-1}^{(m)}+1-\sum_{j}c_{j}^{\mbox{old}}\tau\prod_{m}\hat{g}_{j,t-1}^{(m)}}}.\nonumber 
\end{align}

Therefore, the rate constant for event $v$ is the expected number
of occurrences of this event summed over all times, divided by the
total cross-section of this event also summed over all times.

To summarize, we provide the variational agent-based inference algorithm
below.

\rule{1\columnwidth}{1pt}

\textbf{\uline{Variational Inference with Gillespie Algorithm\hspace*{\fill}}}

Given observations $y_{t}^{(m)}$ for $t=1,\dots,T$ and $m=1,\dots,M$,
and the stochastic kinetic model of a complex system defined by Eq.
\ref{eq:DTSKM}, find $x_{t}^{(m)}$, $v_{t}^{(m)}$ and rate constants
$c_{k}$ for $k=1,\dots,V$. 
\begin{itemize}
\item Latent state inference. Iterate through the following forward pass
and backward pass until convergence, where $P(x_{t}^{(m)},v_{t}|x_{t-1}^{(m)})$
is given by Eq. \ref{eq:marginalEventRate}.

\begin{itemize}
\item Forward pass. For $t=1,\dots,T$ and $m=1,\dots,M$, update $\alpha_{t}^{(m)}(x_{t}^{(m)})$
according to
\begin{align*}
\hspace*{-4em} & \alpha_{t}^{(m)}(x_{t}^{(m)})\leftarrow\frac{1}{Z_{t}}\sum\limits _{x_{t-1}^{(m)},v_{t}\hidewidth}\alpha_{t-1}^{(m)}(x_{t-1}^{(m)})P(x_{t}^{(m)},v_{t}|x_{t-1}^{(m)})P(y_{t}^{(m)}|x_{t}^{(m)}).
\end{align*}

\item Backward pass. For $t=T,\dots,1$ and $m=1,\dots,M$, update $\beta_{t-1}^{(m)}(x_{t-1}^{(m)})$
according to
\begin{align*}
\hspace*{-4em} & \beta_{t-1}^{(m)}(x_{t-1}^{(m)})\leftarrow\frac{1}{Z_{t}}\sum\limits _{v_{t},x_{t}^{(m)}\hidewidth}P(x_{t}^{(m)},v_{t}|x_{t-1}^{(m)})P(y_{t}^{(m)}|x_{t}^{(m)})\beta_{t}^{(m)}(x_{t}^{(m)}).
\end{align*}

\end{itemize}
\item Parameter estimation. Iterate through latent state inference (above)
and rate constants estimate of $c_{k}$ according to Eq. \ref{eq:marginalRateConst},
until convergence. 
\end{itemize}
\rule{1\columnwidth}{1pt}

\section{Experimental Results}

\label{sec:examples}In this section, we evaluate the performance
of variational agent-based inference for two applications: epidemic
dynamics and traffic dynamics. We selected these because they are
important applications with significant practical value.

\subsection{Epidemic Dynamics}

In this section, we infer the progression of common cold in a dynamic
social network using an agent-based susceptible-infectious-susceptible
(SIS) model at the individual level \\
through a small numbr of volunteers who report their symptoms, and
estimate the total number of infectious individuals. Being able to
estimate the outbreak of an epidemic in advance and determine who
has the highest probability of infection is important for health-care
providers and health policy researchers who must optimize limited
medical resources. Conventional agent-based epidemic simulators \cite{eubank2004modelling,hufnagel2004forecast,salathe2010high,isella2011s,edlund2010spatiotemporal}
lack the capability to infer epidemic spreading with symptoms observations
in the social network, and thus the sample paths given by these simulators
can differ significantly from the truth.

In the SIS dynamics, each individual is either infectious ($I$) or
susceptible ($S$), and the system has three events: a) an in infectious
individual in the network infects a susceptible individual and turns
that person infectious in the network with rate constant $c_{1}$
(probability per unit time), b) an infectious individual recovers
and becomes susceptible again with rate constant $c_{2}$, and 3)
a susceptible individual becomes infectious by contacting an infectious
individual from outside the system with rate constant $c_{3}$.

\begin{align*}
 & I+S\to2\times I,\mbox{ infection, rate constant}=c_{1},\\
 & I\to S,\mbox{ recover, rate constant}=c_{2},\\
 & S\to I,\mbox{ infection from outside, rate constant=}c_{3}.
\end{align*}
To model the SIS dynamics at the individual level with Gillespie algorithm,
we assign two ``species'' to each person $p$: $I^{(p)}\in\{0,1\}$,
$S^{(p)}\in\{0,1\}$ and $I^{(p)}+S^{(p)}=1$. The probability for
a susceptible person $p$ to become infectious through one unit time
of contact with an infectious person $q$ is thus $h(x,c_{1})=c_{1}\cdot s^{(p)}\cdot i^{(q)}=c_{1}$.
The mean field probability for the susceptible person $p$ to become
infectious is thus 
\begin{eqnarray*}
\sum\limits _{q\in\mbox{neighbor of }p\hspace*{-3em}}c_{1}\cdot s^{(p)}\cdot\mathbf{E}I^{(q)} & = & c_{1}\cdot s^{(p)}\sum\limits _{q\in\mbox{neighbor of }p\hspace*{-3em}}\mathbf{E}I^{(q)},
\end{eqnarray*}
i.e., the average total number of infectious neighbors in the individual's
social network times the probability of infection per infectious neighbor.
If infection happens, we change $S^{(p)}$ from 1 to 0 and change
$I^{(p)}$ from 0 to 1.

We benchmark the performance of the variational agent-based inference
algorithm using the Dartmouth College campus data set \cite{DaTr07}.
This data set contains the locations of 13,888 on-campus WiFi users
from April 2001 to June 2004. On top of this dynamic social network
we synthesized epidemic progression using the SIS model and set parameters
such that an individual is on average infected twice per year and
takes one week on average to recover. We randomly select 10\% of individuals
as volunteers who report their daily symptoms and from these we infer
the daily infectious/susceptible state of the other 90\% individuals.
As far as we know, there is no real data set with both a large amount
of sensor data and symptom reports; we hope our research encourages
data collection and analysis in this direction.

\begin{figure}[!t]
\centering \subfigure[Detecting infections]{\label{fig:dart_roc}\includegraphics[width=0.49\columnwidth]{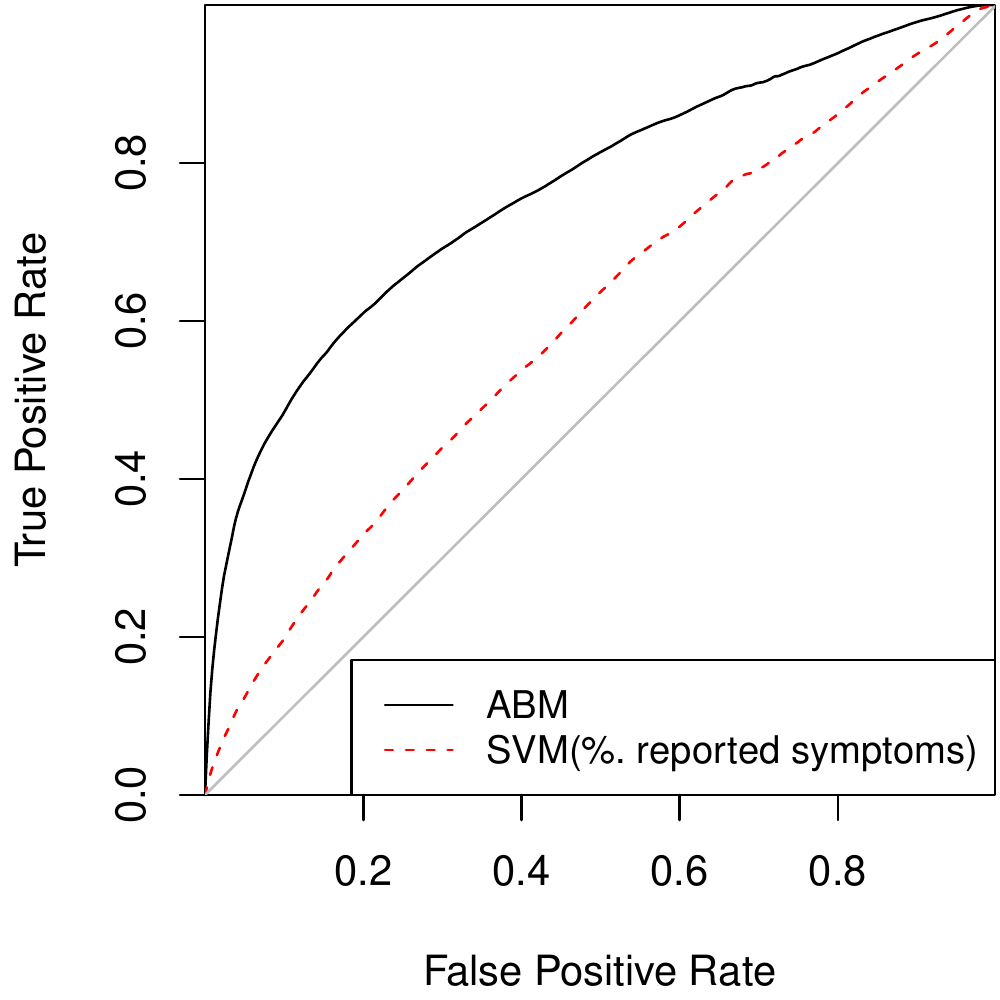}}
\subfigure[\tiny{Estimating infectious population}]{\label{fig:dart_stat}\includegraphics[width=0.49\columnwidth]{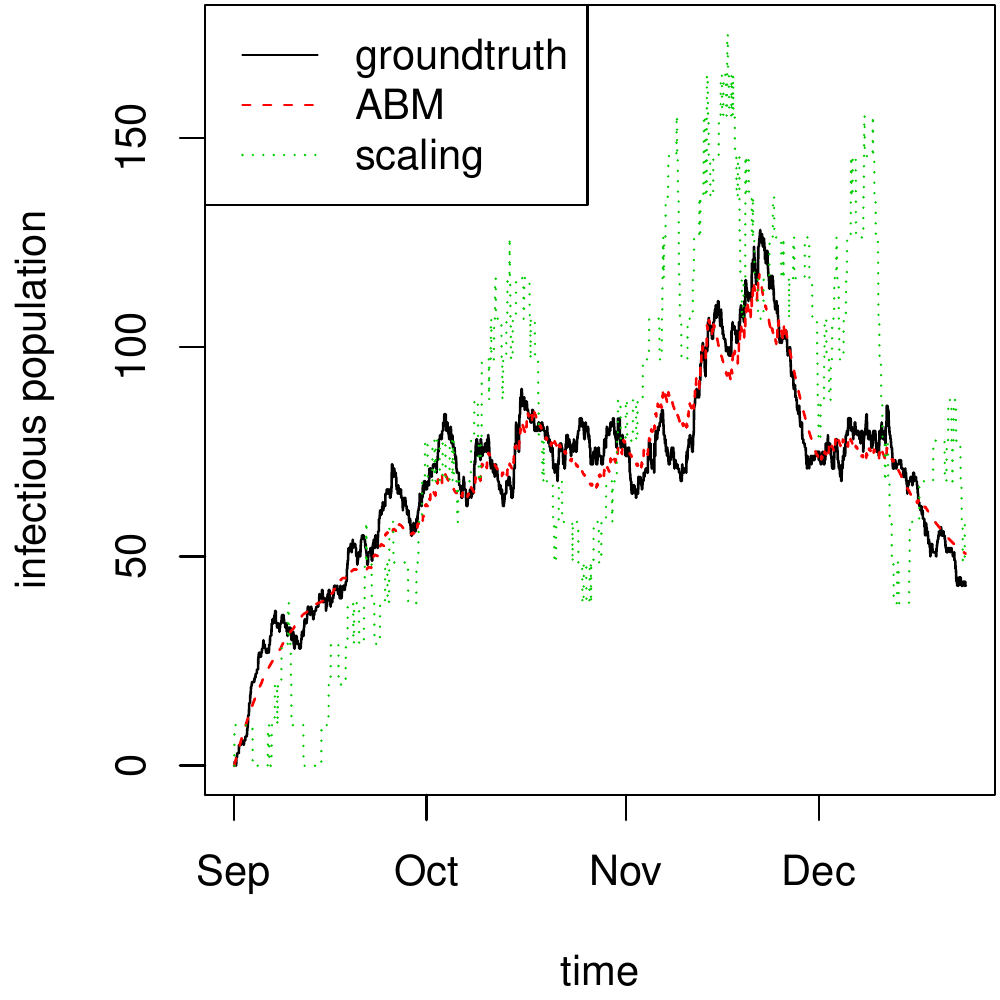}}
\protect\caption{Statistical inference in Dartmouth data}
\label{fig:exp_stat} 
\end{figure}

Figure \ref{fig:dart_roc} compares the receiver operating characteristic
(ROC) curve of predicting whether an individual is infectious using
either a variational agent-based inference algorithm or a support
vector classifier \cite{platt1999probabilistic}. The support vector
classifier (SVC) estimates the probability that an individual is infectious
based on the percentage of his contacts reporting symptoms. Since
only 10\% of individuals are reporting symptoms in the experiment
setup, the likelihood that an individual is infected can be only roughly
estimated by SVC. In addition, infectious individuals make different
contributions to epidemic progression because infectious diseases
from randomly infected individuals go first to the hubs of a social
network then spread to the other nodes from those hubs \cite{Christakis07Obesity},
and SVC has difficulty in capturing such network-related features.
Variational agent-based inference, on the other hand, can correctly
predict 60\% of infections with only a 20\% false positive rate.

Figure \ref{fig:dart_stat} compares the performance in estimating
the number of infectious persons in the 90\% of individuals who do
not report their daily symptoms from daily symptoms reported by the
10\%. A scaling-based method missed the rapid increase of infectious
individuals in early September and overestimated the number of infectious
individuals in mid-October and November. This occurred because not
all infectious individuals contribute to epidemic progression the
same way.

\subsection{Traffic Dynamics}

In this experiment, we predict road traffic up to one hour ahead of
time from a large set of tracked vehicle locations in conjunction
with an agent-based transportation simulator called MATSIM \cite{matsim}.
While tracked vehicle locations from car telematics systems are already
being exploited to provide drivers with real-time traffic information,
the chaotic nature of transportation networks means that an incident
at one location might affect the traffic condition of another location
up to a hundred miles away. A decision made according to current travel
times might therefore be suboptimal, and can in certain cases even
lead to global system breakdown.

Researchers employ a transportation simulator to explain the macroscopic
phenomena of transportation dynamics by simulating how individuals
travel in a real-world transportation network. Such a simulator takes
three primary components as its input: a road network like the one
used in a GPS navigator, a population specification that lists the
location and travel of individuals on a typical day synthesized from
census data and trip surveys with the number of simulated vehicles
matching the number of vehicles in the real world, and a control file
specifying how daily itineraries are scored and how individuals improve
their daily itineraries (and the parameters specifying the modeling
details). From these data, the simulator will proceed to execute travel,
to score daily itineraries, and to perturb daily itineraries in an
attempt to improve them, and then repeat the three steps until equilibrium
is reached. Simulation without continuous observtations about a real-world
transportation network however doesn't tell us whether today's traffic
jams will be formed earlier or last longer.

Many existing algorithms to track and predict real-time traffic dynamics
on the other hand --- vector ARIMA \cite{ben1998dynamit}, state space
models \cite{wang2008real}, neural networks \cite{yin2002urban},
and Bayesian network models \cite{horvitz2012prediction,dong2009network}
--- have difficulty in coping with noisy and missing data, with making
predictions in non-recurrent scenarios, and with explaining predictions
in terms of agent trips. According to Vlahogianni \cite{vlahogianni2014short},
the challenge in short-term traffic forecasting is not only to predict
but also to explain phenomena at the city network level --- to fuse
new data sources such as those from telematics units and to easily
incorporate the effects of non-recurrent conditions. 

To join the event model of a discrete-event simulator with continued
observations about real-world social systems, we make use of the fact
that all discrete event simulators (at least, to the best of our knowledge)
have a way to dump the events happening in a simulation run. As such,
we can reconstruct simulation runs according to the event sequences
and so reconstruct the stochastic discrete-event model from simulation
runs outside the simulator, instead of hacking the source code of
a specific simulator over many man-months. MATSIM, for example, has
about 140 thousand lines of code, and hacking its source code to make
real-time inferences with real-world data wouldn't be easy.

In this way, we dump four events: vehicle leaving a building, vehicle
entering a link, vehicle leaving a link and vehicle entering a building.
From these four events, we have constructed a data frame representing
continued observations of the locations (the link or building) of
all vehicles at equally spaced time steps. From the data frame we
constructed a state transition matrix to represent vehicle dynamics
(with each row and column representing a link/building), along with
entries giving the state transition probabilities according to how
long a vehicle stays at a link/building and how frequently a vehicle
chooses the next link/building. By uniformly sampling a given fraction
of tracked vehicles, we have constructed an observation model that
provides the probability distribution of observed vehicles at a location
given the total number of vehicles there. This system has only one
event $p_{i}\circ l_{j}\to p_{i}\circ l_{k}$, a vehicle $i$ moving
from link/building $j$ to link/building $k$ with rate constant $p_{l_{j},l_{k}}$,
changing the location of vehicle from $X_{t}^{(p_{i})}=l_{j}$ to
$X_{t+1}^{(p_{i})}=l_{k}$, changing the number of vehicles on link
$l_{i}$ from $X_{t}^{(l_{i})}=x_{t}^{(l_{i})}$ to $X_{t+1}^{(l_{i})}=x_{t}^{(l_{i})}-1$,
and changing the number of vehicles on link $l_{k}$ from $X_{t}^{(l_{k})}=x_{t}^{(l_{k})}$
to $X_{t+1}^{(l_{k})}=x_{t}^{(l_{k})}+1$. But the inference is much
more complex due to the interactions between vehicles and links. 

\begin{figure}
\hfill{}\includegraphics[width=0.9\columnwidth]{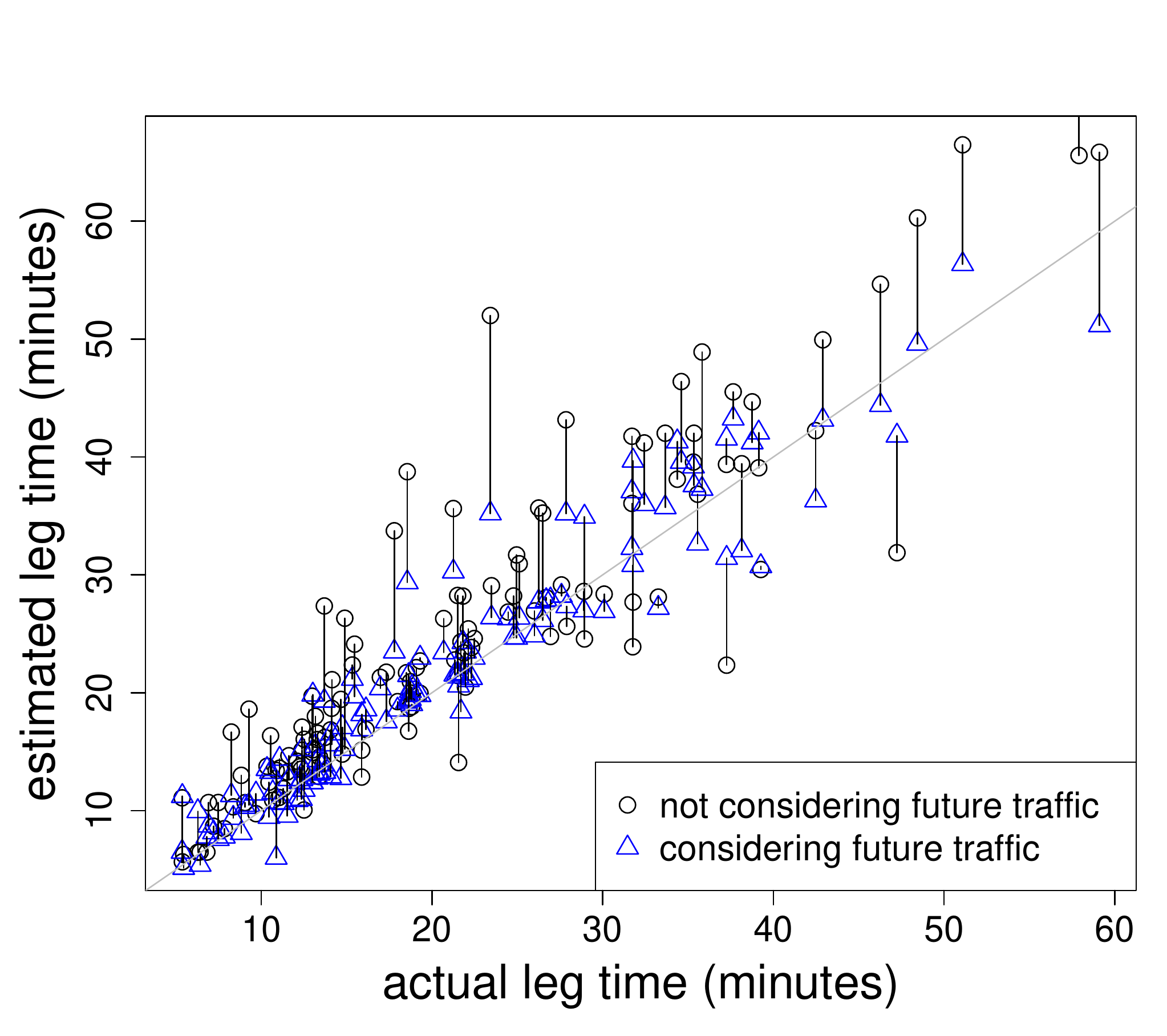}\hfill{}

\protect\protect\caption{\label{fig:TravelTime}Considering future traffic reduces the relative
error of travel time versus actual travel time from 35\% when future
traffic is not considered to 29\% when it is.}
\end{figure}

We employ the mobility traces of more than 500 taxi cabs collected
over 30 days in the San Francisco metropolitan area \cite{epfl-mobility-2009-02-24}
to benchmark the advantages of considering possible future traffic
conditions for individual transportation planning. We extract the
road network from OpenStreetMap, a collaborative project to create
a free editable map of the world. We obtain population distribution
and daily trip statistics from the U.S. Census, and obtain the state
transition matrices from one link/building to another at equilibrium
from simulations. We map the latitudes and longitudes of tracked vehicle
locations to links and buildings because our stochastic inference
is at the link/building level.

To benchmark how the estimation of travel time can be improved by
considering future traffic conditions even though traffic conditions
vary significantly, we extract the 20\% of trips with the highest
difference between actual travel time and estimated travel time according
to traffic at the time of departure, then estimate the average travel
time according to our probabilistic model. Such travel often occurs
at the rising edge of overall traffic volume, during bad weather,
and on less developed roads.

Figure \ref{fig:TravelTime} compares these estimated travel times
according to only traffic at the time of departure with the same data
considering possible future traffic estimations through a random sample
of 128 trips. The travel estimations that do \emph{not} project future
traffic exhibit on average a 35\% relative error (in comparison with
a 29\% when projecting future traffic), and therefore occasionally
differ significantly from the actual travel time.

As such, we can combine the sporadically observed vehicle locations
with the large compilation of typical trip plans to continuously estimate
current and future traffic conditions. Starting from the number and
behavior of tracked vehicles in a road link, we can determine the
total number of vehicles in the link by scaling and estimating traffic
conditions. If we trace the origins and destinations of the estimated
number of vehicles through the factorial stochastic process model
(filling any gaps with prior individual travel behaviors), we can
extract information about the traffic at other road links. If we then
iterate estimations between the traffic at links and the trip choices
of simulated vehicles, we improve our estimation of both.

\section{Conclusions and Discussions}

\label{sec:Conclusions}In this paper, we have developed a variational
method to make inferences about a real-world system from continuous
imperfect observations about the system, using an agent-based model
that describes the dynamics of this system. To demonstrate the value
of combining the power of big data and the power of model-thinking
in the stochastic process framework, we show how we can track epidemics
at the individual level from only a small number of volunteers who
report their symptoms, and we make short-term predictions about road
traffic from sporadically observed probe vehicles. This is just a
taste of what this powerful combination of approaches can do, and
we expect to see further applications and theoretical development
to test the bounds of this methodology. 

\appendix

The duality between Eq. \ref{eq:KL} and Eq. \ref{eq:SKMEvidence}
is a duality between maximum relative entropy and maximum pseudo-likelihood.
To get the dual form of the Bethe variational problem in Eq. \ref{eq:SKMEvidence},
we set the derivatives of Eq. \ref{eq:Lagrange} over $\xi_{t}(x_{t-1,t}v_{t})$
and $\gamma_{t}^{(m)}(x_{t}^{(m)})$ to 0. 

\begin{align*}
 & {\scriptstyle \frac{\partial L}{\partial\xi_{t}(x_{t-1,t},v_{t})}=\log\frac{\xi_{t}(x_{t-1,t},v_{t})}{p(x_{t},v_{t},y_{t}|x_{t-1})}+1-\sum_{m}\beta_{t,x_{t}^{(m)}}^{(m)}-\sum\limits _{m}\alpha_{t-1,x_{t-1}^{(m)}}^{(m)}\stackrel{\mbox{set}}{=}0}\\
 & \Rightarrow{\scriptstyle \xi_{t}(x_{t-1,t},v_{t})\propto\exp\left(\sum\limits _{m}\alpha_{t-1,x_{t-1}^{(m)}}^{(m)}\right)p(x_{t},v_{t},y_{t}|x_{t-1})\exp\left(\sum\limits _{m}\beta_{t,x_{t}^{(m)}}^{(m)}\right),}\\
 & \frac{\partial L}{\partial\gamma_{t}^{(m)}(x_{t}^{(m)})}=\log\gamma_{t}^{(m)}(x_{t}^{(m)})+1-\beta_{t,x_{t}^{(m)}}^{(m)}-\alpha_{t,x_{t}^{(m)}}^{(m)}\stackrel{\mbox{set}}{=}0\\
 & \Rightarrow\gamma_{t}^{(m)}(x_{t}^{(m)})\propto\exp\left(\alpha_{t,x_{t}^{(m)}}^{(m)}+\beta_{t,x_{t}^{(m)}}^{(m)}\right).
\end{align*}

After taking $\xi_{t}(x_{t-1,t},v_{t})$ and $\gamma_{t}^{(m)}(x_{t}^{(m)})$
into Eq \ref{eq:KL}, we get Eq. \ref{eq:SKMEvidence}.

To derive Eq. \ref{eq:marginal2Slice}, we marginalize Eq. \ref{eq:SKM2Slice}
over all $x_{t-1,t}^{(m')}$ for $m'\ne m$.

\begin{align}
 & \xi(x_{t-1}^{(m)},v_{t},x_{t}^{(m)})\nonumber \\
 & ={\scriptstyle \frac{1}{Z_{t}}\sum\limits _{x_{t-1,t},v_{t}:\mbox{fix }x_{t-1,t}^{(m)},v_{t}\hidewidth}p(x_{t},v_{t}|x_{t-1})\prod_{m}\alpha^{(m)}(x_{t-1}^{(m)})\beta^{(m)}(x_{t}^{(m)})p^{(m)}(y_{t}^{(m)}|x_{t}^{(m)})}\nonumber \\
 & \propto{\scriptstyle c_{k}\tau\prod\limits _{\hidewidth m'\ne m}\sum\limits _{x_{t-1,t}^{(m')}\hidewidth}\alpha_{t-1}^{(m')}g^{(m')}(x_{t-1}^{(m')})\beta_{t}^{(m')}p(y_{t}^{(m')}|x_{t}^{(m')})1(\Delta x_{t}^{(m')}=\Delta_{k}^{(m')})}\nonumber \\
 & {\scriptstyle \cdot\alpha_{t-1}^{(m)}(x_{t-1}^{(m)})g^{(m)}(x_{t-1}^{(m)})\beta_{t}^{(m)}(x_{t}^{(m)})p(y_{t}^{(m)}|x_{t}^{(m)})1(\Delta x_{t}^{(m)}=\Delta_{k}^{(m)})}\nonumber \\
 & \hspace*{15em}\mbox{ if }v_{t}=k\ne\emptyset,\label{eq:marginal:xi_k}\\
 & \propto{\scriptstyle \prod\limits _{\hidewidth m'\ne m}\sum\limits _{x_{t-1,t}^{(m')}\hidewidth}\alpha_{t-1}^{(m')}(x_{t-1}^{(m')})\beta_{t}^{(m')}(x_{t}^{(m')})p(y_{t}^{(m')}|x_{t}^{(m')})1(\Delta x_{t}^{(m')}=0)}\nonumber \\
 & {\scriptstyle \cdot\alpha_{t}^{(m)}(x_{t-1}^{(m)})\beta_{t}^{(m)}(x_{t}^{(m)})p(y_{t}^{(m)}|x_{t}^{(m)})1(\Delta x_{t}^{(m)}=0)}\nonumber \\
 & {\scriptstyle -\sum\limits _{k}\ c_{k}\tau\prod\limits _{\hidewidth m'\ne m}\sum\limits _{x_{t-1,t}^{(m')}\hidewidth}\alpha_{t-1}^{(m')}g^{(m')}(x_{t-1}^{(m')})\beta_{t}^{(m')}p(y_{t}^{(m')}|x_{t}^{(m')})1(\Delta x_{t}^{(m')}=0)}\nonumber \\
 & {\scriptstyle \cdot\alpha_{t-1}^{(m)}(x_{t-1}^{(m)})g^{(m)}(x_{t-1}^{(m)})\beta_{t}^{(m)}(x_{t}^{(m)})p(y_{t}^{(m)}|x_{t}^{(m)})1(\Delta x_{t}^{(m)}=0)}\nonumber \\
 & \hspace*{15em}\mbox{ if }v_{t}=\emptyset.\label{eq:margina:xi_0}
\end{align}
After deviding Eq. \ref{eq:marginal:xi_k} and Eq. \ref{eq:margina:xi_0}
by 
\[
{\scriptstyle \prod_{m'\ne m}\sum\limits _{x_{t-1,t}^{(m')}\hidewidth}\alpha_{t-1}^{(m')}(x_{t-1}^{(m')})\beta_{t}^{(m')}(x_{t}^{(m')})p(y_{t}^{(m')}|x_{t}^{(m')})1(\Delta x_{t}^{(m')}=0)}
\]
 we get Eq. \ref{eq:marginal2Slice}, $\tilde{g}_{k,t-1}^{(m)}$,
$\hat{g}_{k,t-1}^{(m)}$ and $Z_{t}$.

\balance\bibliographystyle{plain}
\bibliography{abm}

\end{document}